\documentclass{aa}

\usepackage{graphicx}

\usepackage[varg]{txfonts}
\usepackage{natbib}
\bibpunct{(}{)}{;}{a}{}{,}
\usepackage{array}
\usepackage{xspace}
\usepackage{lscape}
\usepackage{url}
\usepackage{arydshln}
\usepackage[hyperfootnotes=false, linktocpage=true, breaklinks=true, colorlinks=true, linkcolor=blue, citecolor=blue, urlcolor=blue]{hyperref}
\usepackage[all]{hypcap}
\usepackage{longtable}
\usepackage{afterpage}

\usepackage[dvipsnames]{xcolor}

\newcommand{\FeKa}{Fe K\ensuremath{\alpha}\xspace}

\newcommand{\kms}{\ensuremath{\mathrm{km\ s^{-1}}}\xspace}
\newcommand{\NH}{\ensuremath{N_{\mathrm{H}}}\xspace}

\newcommand{\xmm}{{\it XMM-Newton}\xspace}
\newcommand{\chandra}{{\it Chandra}\xspace}
\newcommand{\swift}{{\it Swift}\xspace}

\newcommand{\spitzer}{{\it Spitzer}\xspace}
\newcommand{\hst}{{HST}\xspace}

\newcommand{\eso}{{ESO~113-G010}\xspace}
\newcommand{\Ha}{H\ensuremath{\alpha}\xspace}
\newcommand{\Hb}{H\ensuremath{\beta}\xspace}

\newcommand{\herschel}{{\it Herschel}\xspace}
\newcommand{\ic}{{IC~4329A}\xspace}

\newcommand{\nustar}{{\it NuSTAR}\xspace}
\newcommand{\suzaku}{{\it Suzaku}\xspace}

\newcommand{\athena}{{\it Athena}\xspace}

\newcommand{\ergcm}{{\ensuremath{\rm{erg\ cm}^{-2}\ \rm{s}^{-1}\ {\AA}^{-1}}}\xspace}
\newcommand{\ergflux}{{\ensuremath{\rm{erg\ cm}^{-2}\ \rm{s}^{-1}}}\xspace}
\newcommand{\ergs}{{\ensuremath{\rm{erg\ s}^{-1}}}\xspace}
\newcommand{\cm}{{\ensuremath{\rm{cm}^{-2}}}\xspace}

\newcommand{\spex}{\xspace{\tt SPEX}\xspace}
\newcommand{\pion}{\xspace{\tt pion}\xspace}

\newcommand{\micron}{\ensuremath{\mu{\mathrm{m}}}\xspace}

\newcommand{\ebv}{\ensuremath{{E(B-V)}}\xspace}
\newcommand{\MBH}{\ensuremath{{M_{\rm BH}}}\xspace}

\mathchardef\mhyphen="2D

\begin{document}

\title{Probing the nature and origin of dust in the reddened quasar IC~4329A with global modelling from X-ray to infrared}

\author{
Missagh Mehdipour
\and
Elisa Costantini
}
\institute{
SRON Netherlands Institute for Space Research, Sorbonnelaan 2, 3584 CA Utrecht, the Netherlands\\ \email{M.Mehdipour@sron.nl}
}
\date{Received 24 June 2018 / Accepted 13 August 2018}

\abstract
{
Cosmic dust is a key tracer of structure formation and evolution in the universe. In active galactic nuclei (AGN) the origin and role of dust are uncertain. Here, we have studied dust in the X-ray bright and reddened type-1 quasar \ic, which exhibits an ionised AGN wind. We incorporated high-resolution X-ray and mid-IR spectroscopy, combined with broad-band continuum modelling, to investigate the properties of dust in this AGN. We used new \chandra HETGS observations taken in June 2017, as well as archival data from \xmm, \swift, \hst, \spitzer, IRAS, and \herschel for our IR-optical-UV-X-ray modelling. Two distinct components of dust in \ic are found. One component is in the interstellar medium (ISM) of the host galaxy, and the other is a nuclear component in the AGN torus and its associated wind. The emitting dust in the torus is evident in mid-IR emission (9.7 and 18 \micron features), while dust in the wind is present through both reddening and X-ray absorption (O, Si, and Fe edge features). The gas depletion factors into dust for O, Si, and Fe are measured. We derive an intrinsic reddening $\ebv \approx 1.0$, which is most consistent with a grey (flat) extinction law. The AGN wind consists of three ionisation components. From analysis of long-term changes in the wind, we determine limits on the location of the wind components. The two lowest ionisation components are likely carriers of dust from the AGN torus. We find that the dust in the nuclear component of \ic is different from dust in the Milky Way. The dust grains in the AGN torus and wind are likely larger than the standard Galactic dust, and are in a porous composite form (containing amorphous silicate with iron and oxygen). This can be a consequence of grain coagulation in the dense nuclear environment of the AGN.
}
\authorrunning{M. Mehdipour \& E. Costantini}
\titlerunning{Nature and origin of dust in IC 4329A}
\maketitle

\section{Introduction}

Cosmic dust is widespread in the universe. It can be used to trace the evolutionary paths of planets, stars, and even black holes. Dust grains in the ISM of galaxies alter the appearance of the observable universe through absorption and scattering of photons. Thus, determining the effects of dust is needed for studying a wide range of astrophysical phenomena. Indeed a significant fraction of ISM is locked up in dust grains (e.g. \citealt{Jenk09}). However, many aspects of cosmic dust remain poorly understood, such as: their chemical composition and origin; their physical properties and spectral signatures; their formation and evolution, and destruction mechanisms; and their role and impact on their environment. In AGN the properties of dust are particularly uncertain (see e.g. the review by \citealt{Li07}). Also, dust in AGN can be associated to different possible origins, such as: dust lanes of recent galaxy merger remnants, or dusty winds from the AGN (see e.g. \citealt{Komo97,Reyn97,Lee01,Cren01b}).

Dust is an essential player in the unification theory of AGN, where the supermassive black hole (SMBH) and the accretion disk are surrounded by an optically-thick dusty torus. The observational properties of AGN are hence strongly influenced by our viewing angle relative to the orientation of this obscuring torus (\citealt{Anto85,Anto93,Urry95}). Yet, our knowledge of dust properties in the AGN torus and its environment is limited. 

Accretion onto SMBHs at the core of AGN is accompanied by winds of gas, which couple the SMBHs to their environment. The observed relations between SMBHs and their host galaxies, such as the M--$\sigma$ relation \citep{Ferr00}, indicate that SMBHs and their host galaxies are likely co-evolved through a feedback mechanism. The AGN winds can play a key role in this co-evolution as they can significantly impact star formation (e.g. \citealt{Silk98}) and chemical enrichment of the surrounding intergalactic medium (e.g. \citealt{Oppen06}). However, there are significant gaps in our understanding of the outflow phenomenon in AGN, which cause major uncertainties in determining their role and impact in galaxy evolution.

The origin and physical structure of AGN winds are generally poorly understood. Different mechanisms have been postulated for the launch and driving of winds from either the accretion disk or the AGN torus (e.g. \citealt{Krol01,Prog04,Fuku10}). However, their association to the different kinds of winds found from observations is uncertain. AGN winds can originate as either thermally-driven \citep{Krol01} or radiatively-driven \citep{Doro08} winds from the dusty torus. Indeed the warm-absorber winds, which are commonly detected in the UV and X-ray spectra of bright AGN, are most consistent with being torus winds (e.g. \citealt{Kaas12,Meh18}); thus dust could be mixed with such winds. Importantly, the infrared (IR) radiation pressure on dust grains can boost winds from the torus \citep{Doro11}. Therefore, establishing the existence of dust in AGN winds is important for understanding the driving mechanism of AGN winds, and defining their impact on their environment. This is needed for assessing the contribution of such winds to AGN feedback. There are observational evidence, which indicate that AGN winds are likely carriers of dust into the ISM. For example, in the case of \object{ESO~113-G010} \citep{Meh12}, presence of dust embedded in the AGN wind was inferred based on the altering of the AGN emission, dust-to-gas ratio arguments, and the properties of the wind.

In order to advance our understanding of cosmic dust in the universe we need to utilise all available tools at our disposal. The X-ray energy band, which is invaluable for exploring both the cold and hot gas, is a relatively new scientific window for dust studies. The X-ray absorption fine structures (XAFS) at the K edge of O, Mg, Si, Fe, and the LII and LIII edges of Fe provide distinct and unblended signatures of dust grains in X-rays (e.g. \citealt{Lee09,Cost12,Zeeg17,Roga18}). Therefore, in addition to the traditional low-energy domain observations, like in the IR, the X-ray band enables us to directly access the chemical composition of dust in the diffuse ISM of galaxies. Thus, high-resolution X-ray spectroscopy provides a powerful and sensitive diagnostic tool to probe the properties of both gas and dust. This is invaluable for understanding the formation and evolution history of galaxies, including the host galaxies of AGN. Indeed X-ray spectroscopy can help in studying the chemistry of dust in AGN (e.g. \citealt{Lee13}), which is an important indicator of the evolutionary phase of AGN.

\begin{figure}[!tbp]
\centering
\hspace{-0.8cm}\resizebox{0.95\hsize}{!}{\includegraphics[angle=0]{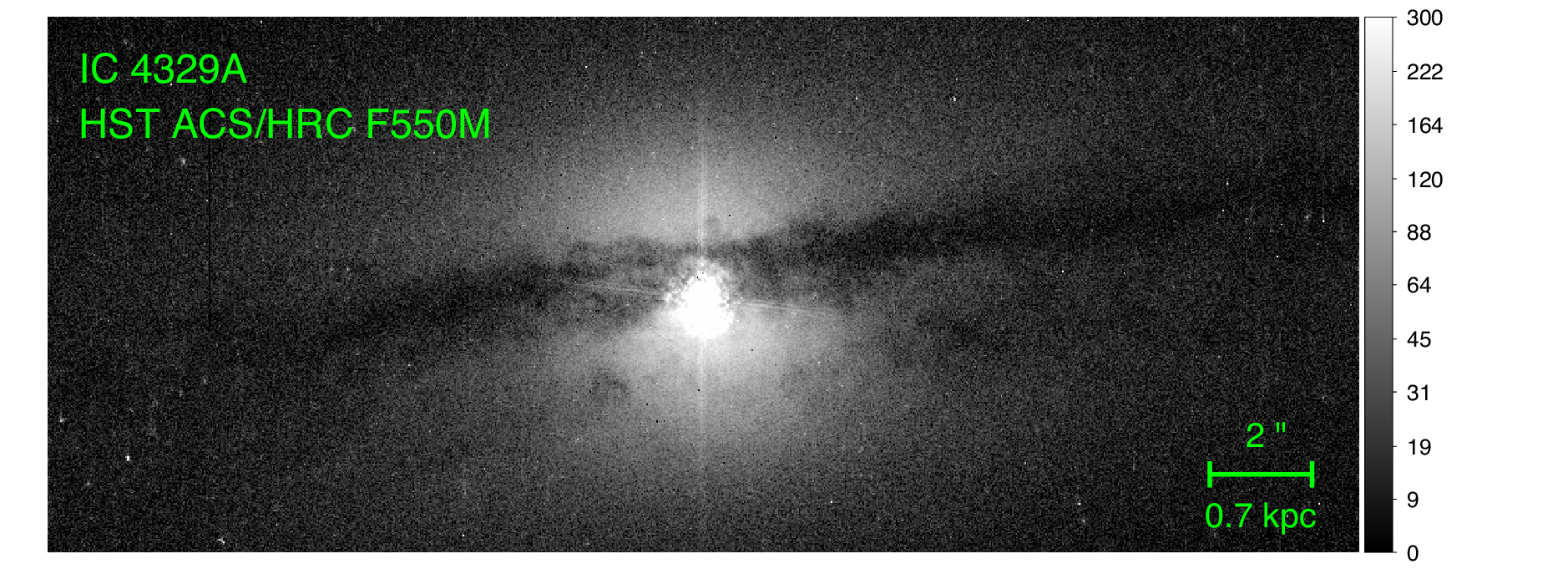}\hspace{-2.8cm}}
\caption{Nuclear region of \ic as observed by the HST, showing the presence of a dust lane covering the nucleus. The image is obtained from an observation taken with the ACS/HRC F550M filter on 22 February 2006, and is displayed with a logarithmic intensity scale.}
\label{dustlane_fig}
\end{figure}

\object{IC~4329A} is bright nearby AGN at redshift ${z = 0.016054}$ \citep{Will91}. It has been described as `an extreme Seyfert galaxy' \citep{Disn73} and `the nearest quasar' \citep{Wils79}, based on the spectroscopy of its broad optical emission lines. However, it is technically classified as a Seyfert 1.2 by \citet{Vero06}. The host galaxy of the AGN is highly inclined (i.e. edge-on), with the observed ratio of minor to major axis ${b/a = 0.28}$ \citep{deVa91}. A prominent dust lane bisects the nucleus of \ic, which can be seen in the HST image of Fig. \ref{dustlane_fig}. The dust lane is indicative of the past merger history of this galaxy. \ic is a member of a group of seven galaxies \citep{Koll89}. It is displaced by a projected distance of 59 kpc from the giant lenticular galaxy \object{IC 4329} \citep{Wols95}. The relative orientations of the axes of the AGN and the disk of the host galaxy may have been influenced by interaction between \ic and its massive neighbour IC~4329 \citep{Wols95}. 

The mass of the supermassive black hole (\MBH) in \ic is not accurately determined. The \MBH from reverberation study is poorly constrained due to low quality lightcurves, and hence unreliable lag measurements, from which \citet{Pet04} estimate an upper limit of $\sim 3 \times 10^{7}$~$M_{\odot}$. However, measurements using other methods find higher \MBH. Using the empirical relation derived by \citet{McHa06} between the bolometric luminosity $L_{\rm bol}$, \MBH, and the break frequency in the X-ray power spectral density function (PSD), \citet{Mark09} calculate $M_{\rm BH} = 1.3_{-0.3}^{+1.0} \times 10^{8}$~$M_{\odot}$. Also, using relations between \MBH and the stellar velocity dispersion $\sigma_*$, \citet{Mark09} find $\MBH \approx 2_{-1}^{+2} \times 10^{8}$~$M_{\odot}$. Furthermore, \citet{deLa10} report $\MBH \sim 1.2 \times 10^{8}$~$M_{\odot}$. From the AGN sample study of \citet{Vasu10}, the black hole mass is estimated from the $K$-band luminosity of the host galaxy bulge, which in the case of \ic is found to be about $2 \times 10^{8}$~$M_{\odot}$. Therefore, in our calculations we assume \MBH is $\sim$1--2~$\times 10^{8}$~$M_{\odot}$.

\citet{Stee05} studied the X-ray absorption in \ic with \xmm RGS spectroscopy. They found absorption by neutral gas in the host galaxy of the AGN, as well as warm absorption by an AGN wind. Despite significant absorption, \ic is bright enough for high-resolution X-ray spectroscopy, which is often not the case for many reddened and absorbed AGN, making \ic a valuable target. The column density \NH of the neutral gas was measured to be about ${1.7 \times 10^{21}}$~\cm \citep{Stee05}. The warm absorber was found to have four ionisation components, with a total \NH of about ${1.0 \times 10^{22}}$~\cm. The velocity of the warm absorber components ranges from ${-200 \pm 100}$ to ${+20 \pm 160}$~\kms. Possible X-ray spectral features of dust were not investigated in \citet{Stee05}.

Dust reddened and absorbed AGN are often too faint for dust X-ray spectroscopy because of strong absorption. Therefore, finding a suitable target is key. For our X-ray spectroscopic investigation of dust in winds, the AGN target must meet these required selection criteria: (1) bright enough in X-rays; (2) displays significant intrinsic reddening; (3) exhibits evidence of dust features in both X-rays and IR; (4) shows the presence of an AGN wind; (5) minimum amount of reddening and absorption contamination by the Milky Way in our of sight. Among many candidates that we systematically searched, \ic is one of the most suitable AGN that meets all these criteria. Our newly acquired \chandra-HETGS observations of \ic are presented for the first time in this paper.

The structure of the paper is as follows. The observations and the reduction of data are described in Sect. \ref{data_sect}. The modelling of the spectral energy distribution (SED) is presented in Sect. \ref{sed_sect}. The modelling of the X-ray absorption by the ISM gas and the AGN wind are explained in Sect. \ref{wind_sect}. The multi-wavelength analysis of dust in \ic based on reddening, IR emission features, and X-ray absorption features, is presented in Sect. \ref{dust_sect}. We discuss all our findings in Sect. \ref{discussion}, and give concluding remarks in Sect. \ref{conclusions}.

The spectral analysis and modelling presented in this paper were done using the {\tt SPEX} package \citep{Kaa96} v3.04.00. The spectra shown in this paper are background-subtracted and are displayed in the observed frame. We use C-statistics for spectral fitting and give errors at the $1\sigma$ confidence level. We adopt a luminosity distance of 69.61 Mpc in our calculations with the cosmological parameters ${H_{0}=70\ \mathrm{km\ s^{-1}\ Mpc^{-1}}}$, $\Omega_{\Lambda}=0.70$, and $\Omega_{m}=0.30$. We assume proto-solar abundances of \citet{Lod09} in all our computations in this paper.

%

\section{Observations and data processing}
\label{data_sect}

The observation log of data used in our spectral analysis are provided in Table \ref{log_table}. We describe the processing of data from different observatories in the following.

\subsection{X-ray data}

All \chandra observations of \ic have been taken with HETGS \citep{Cani05}. In all the HETGS observations, the ACIS camera was operated in the timed exposure (TE) read mode and the faint data mode. The data were reduced using the Chandra Interactive Analysis of Observations ({\tt CIAO}, \citealt{Frus06}) v4.9 software and the calibration database (CALDB) v4.7.3. The {\tt chandra\_repro} script of {\tt CIAO} and its associated tools were used for the reduction of the data and production of the final grating products (PHA2 spectra, RMF and ARF response matrices). The MEG and HEG +/- first-order spectra and their response matrices were combined using the CIAO {\tt combine\_grating\_spectra} script. The short-term X-ray variability, seen by the HETGS observations on days timescale, correspond to about 10\% flux variation around the mean. This is a small variability, allowing us to stack the individual spectra in order to enhance the signal-to-noise ratio. The fitted spectral range is 2.5--26~\AA\ for MEG, and 1.55--14.5~\AA\ for HEG. Over these energy bands, the HEG\,/\,MEG flux ratio is nearly constant at 0.952. We take into account this instrumental flux difference between HEG and MEG by re-scaling the normalisation of HEG relative to MEG in our spectral modelling. The low and high energy data outside of these ranges are ignored because of deviations in the HEG\,/\,MEG flux ratio caused by the increasing calibration uncertainties of the instruments.

The \xmm data were processed using the Science Analysis System (SAS v16.0.0). The RGS \citep{denH01} instruments were operated in the standard Spectro+Q mode for the \xmm observations of \ic (Table \ref{log_table}). The data were processed through the {\tt rgsproc} pipeline task; the source and background spectra were extracted and the response matrices were generated. We filtered out time intervals with background count rates $> 0.1\ \mathrm{count\ s}^{-1}$ in CCD number 9. The {\tt rgscombine} task was used to stack the RGS first-order spectra. Our fitted spectral range is 7--35~\AA\ for RGS. The EPIC-pn instrument \citep{Stru01} was operated in the Full-Frame mode with the Medium Filter during the 2001 observation, and in the Small-Window mode with the Thin Filter during the 2003 observation. Periods of high flaring background for EPIC-pn (exceeding 0.4 $\mathrm{count\ s}^{-1}$) were filtered out by applying the {\tt \#XMMEA\_EP} filtering. The EPIC-pn spectra were extracted from a circular region centred on the source with a radius of $40''$. The background was extracted from a nearby source-free region of radius $40''$ on the same CCD as the source. The pileup was evaluated to be small at about 2\%. The single and double events were selected for the EPIC-pn ({\tt PATTERN <= 4}). Response matrices were generated for the spectrum of each observation using the {\tt rmfgen} and {\tt arfgen} tasks. The {\tt epicspeccombine} task was used for stacking the EPIC-pn spectra. Our fitted spectral range for EPIC-pn is from 1.38 keV (9 \AA) to 10 keV, since the soft X-ray band (0.35--1.8 keV) is simultaneously modelled with RGS. The EPIC-pn\,/\,RGS flux ratio at the overlapping energy band is found to be 0.91, which we take into account by re-scaling in our simultaneous fitting of the RGS and EPIC-pn spectra. In our spectral modelling we first derived a time-averaged model fitted to the stacked spectra from all available observations, and then applied this model to determine the variability in the X-ray continuum and the X-ray absorption between the 2003 and 2017 epochs, when the deepest X-ray observations were taken.

%
\begin{table}[!tbp]
\begin{minipage}[t]{\hsize}
\setlength{\extrarowheight}{3pt}
\caption{Observation log of the data used in our spectral modelling.}
\centering
\footnotesize
\renewcommand{\footnoterule}{}
\begin{tabular}{l | c c c}
\hline \hline
 & & Obs. date & Length  \\
Observatory & Obs. ID & yyyy-mm-dd & (ks)  \\ 
\hline
\chandra/HETGS & 2177 &  2001-08-26 & 59.1	\\
\chandra/HETGS & 20070 & 2017-06-06 & 91.8	\\
\chandra/HETGS & 19744 & 2017-06-12 & 12.4	\\
\chandra/HETGS & 20095 & 2017-06-13 & 33.3	\\
\chandra/HETGS & 20096 & 2017-06-14 & 19.8	\\
\chandra/HETGS & 20097 & 2017-06-17 & 16.8	\\
\hline
\xmm & 0101040401	& 2001-01-31 & 13.9 \\
\xmm & 0147440101	& 2003-08-06 & 136.0 \\
\hline
\swift/UVOT & 00033058001	& 2013-12-21 & 1.9 \\
\hline
HST/WFPC2/F814W & U5GU0401R	& 2000-02-25 & 0.012 \\
HST/NICMOS/F160W & N4JQ06010 & 1998-05-21 & 0.22 \\
HST/NICMOS/F196N & N4JQ06070 & 1998-05-21 & 0.51 \\
HST/NICMOS/F200N & N4JQ060A0 & 1998-05-21 & 0.90 \\
HST/NICMOS/F222M & N4JQ06040 & 1998-05-21 & 0.26 \\
\hline
\spitzer/IRAC & 12472576 & 2005-07-18 & 0.047  \\
\spitzer/IRAC & 18038784 & 2006-08-13 & 1.3 \\
\spitzer/MIPS & 10642176 & 2006-02-15 & 0.53 \\
\spitzer/IRS & 4848640 & 2004-07-13 & 1.3 \\
\spitzer/IRS & 18506496 & 2007-07-29 & 0.96 \\
\hline
\herschel/PACS/70-\micron & 1342236918 & 2012-01-07 & 0.052\\
\herschel/PACS/160-\micron & 1342236919 & 2012-01-07 & 0.052 \\
\herschel/SPIRE & 1342236198 & 2012-01-02 & 0.169 \\
\hline
\end{tabular}
\end{minipage}
\tablefoot{
The dates correspond to the start time of the observations in UTC. In addition to the above data, we also use fluxes at 60 and 100 $\mu$m from the IRAS survey observations.
}
\label{log_table}
\end{table}

\subsection{Optical and UV data}

In order to determine the optical and UV part of the SED, we made use of photometric data from \xmm OM \citep{Mas01} and \swift UVOT \citep{Romi05}. We use data taken with the V, B, U, UVW1, UVM2 and UVW2 filters. For a description of the reduction of OM and UVOT data, we refer to Appendix A in \citet{Meh15a} and references therein, which applies to the data used here. The size of the circular aperture used for our photometry was set to a diameter of 12$\arcsec$ for OM, and 10$\arcsec$ for UVOT, which is the optimum aperture size based on the calibration of these instruments.

The continuum at the UV and optical, and lower energies, is approximated to be constant over time in our SED modelling. The optical and UV fluxes from the OM and UVOT are consistent with each other at the overlapping filters despite being taken years apart. Furthermore, from examining the long-term variability in the UVOT observations, we find that the flux variation in the V band is about 4\% around the mean. Therefore, the continuum at the optical and lower energies is not too variable for the purpose of approximating the SED by simultaneously fitting the individual observations given in Table \ref{log_table}.

\subsection{Infrared data}

To determine the SED spanning near-IR to far-IR energies, we have utilised data from \hst, \spitzer, IRAS, and \herschel. The \hst images from Wide Field and Planetary Camera 2 (WFPC2) F814W filter, and the Near Infrared Camera and Multi-Object Spectrometer (NICMOS) F160W, F196N, F200N, F222M filters, were retrieved from Mikulski Archive for Space Telescopes (MAST). We carried out aperture photometry with a diameter of 10\arcsec on the central source. 

The \spitzer Multi-band Imaging Photometer (MIPS) spectrum was retrieved from the Spitzer Heritage Archive (SHA). The MIPS instrument, operating in the SED mode, provides a low-resolution spectrum from 53 to 100 \micron. Furthermore, the \spitzer Infrared Array Camera (IRAC) photometric measurements at 3.6, 4.5, 5.8, 8.0 \micron were taken from the \spitzer study of a sample of AGN by \citet{Gall10}. The \spitzer Infrared Spectrograph (IRS) observations, operating in the IRS Stare mode, were used to extract the spectra. The IRS spectra from the low-resolution modules, short-low (SL) and long-low (LL), were retrieved from the Combined Atlas of Sources with Spitzer IRS Spectra (CASSIS). The spectra from the high-resolution modules, short-high (SH) and long-high (LH), were processed through the c2d pipeline and optimal PSF extraction.   

The Infrared Astronomical Satellite (IRAS) flux measurement at 60 and 100 $\mu$m were obtained from the IRAS Faint Source catalogue v2.0 \citep{Mosh90}. We have also utilised far-infrared (FIR) data from \herschel PACS (70 and 160 \micron filters) and SPIRE (250, 350, and 500 \micron filters). The photometric flux measurements were obtained from the \herschel-\swift/BAT AGN sample studies of \citet{Mele14} and \citet{Shim16}, which report on the PACS and SPIRE observations, respectively.

\section{Determination of the spectral energy distribution}
\label{sed_sect}

Here we present our modelling of the components of the IR-optical-UV-X-ray continuum. We note that our continuum modelling is tied to the modelling of the X-ray absorption, reddening, and dust IR emission features, which are presented separately in later sections (Sects. \ref{wind_sect} and \ref{dust_sect}).

\subsection{Primary IR-optical-UV-X-ray continuum of the AGN}
\label{continuum_sect}

We started by applying a power-law component ({\tt pow}) to fit the X-ray spectrum of \ic (Fig. \ref{overview_fig}). The X-ray spectrum is strongly absorbed, and the modelling of this absorption is described in Sect. \ref{wind_sect}. The X-ray power-law continuum model represents Compton up-scattering of the disk photons in an optically-thin and hot corona in \ic. The high-energy exponential cut-off of the power-law was set to 186~keV \citep{Bren14} based on \nustar and \suzaku observations. A low-energy exponential cut-off was also applied to the power-law continuum to prevent exceeding the energy of the seed disk photons. The photon index $\Gamma$ of the power-law is ${\Gamma = 1.78 \pm 0.01}$.

\begin{figure}[!tbp]
\centering
\resizebox{1.043\hsize}{!}{\hspace{-0.85cm}\includegraphics[angle=0]{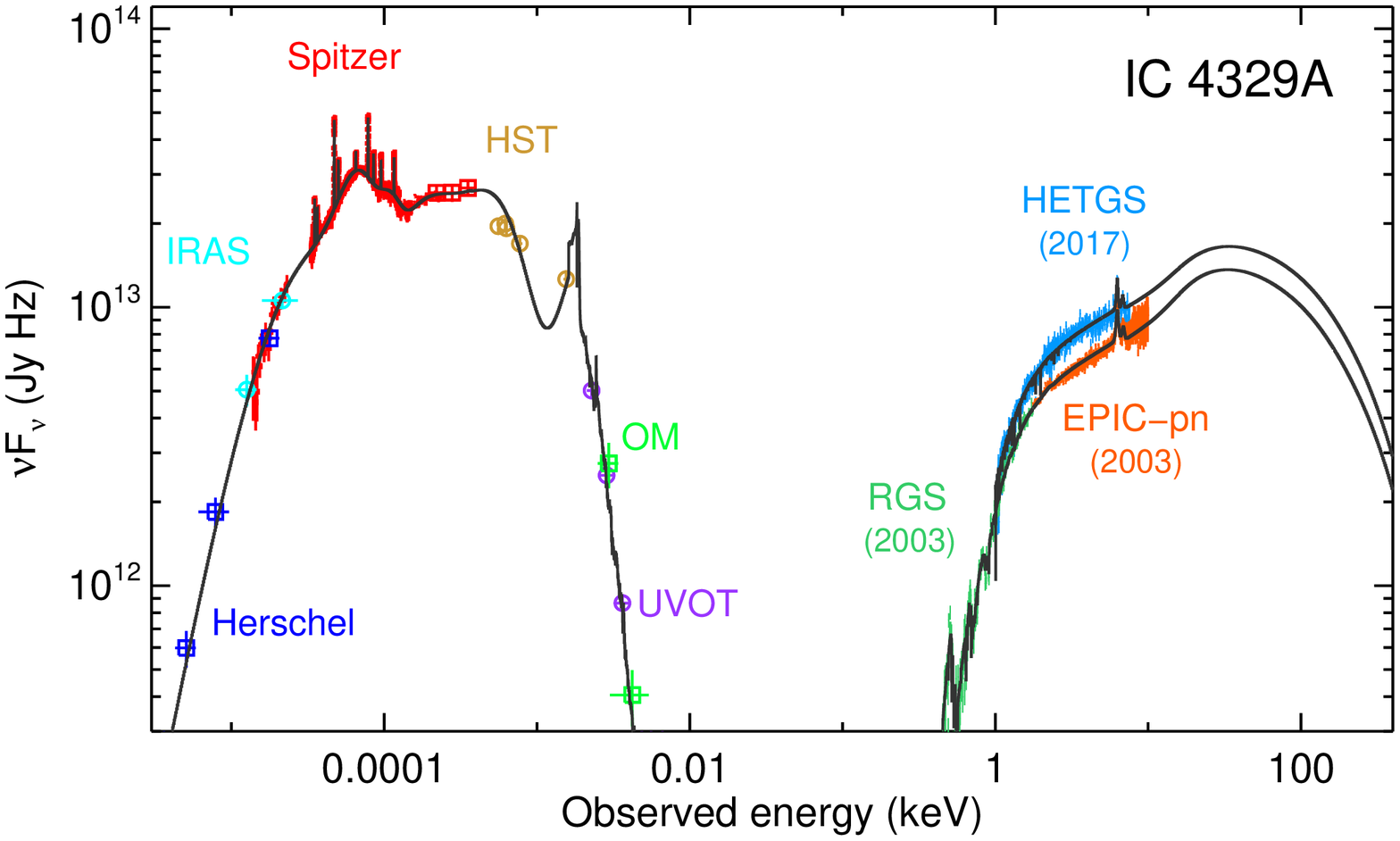}}\vspace{-0.3cm}
\resizebox{1.043\hsize}{!}{\hspace{-0.85cm}\includegraphics[angle=0]{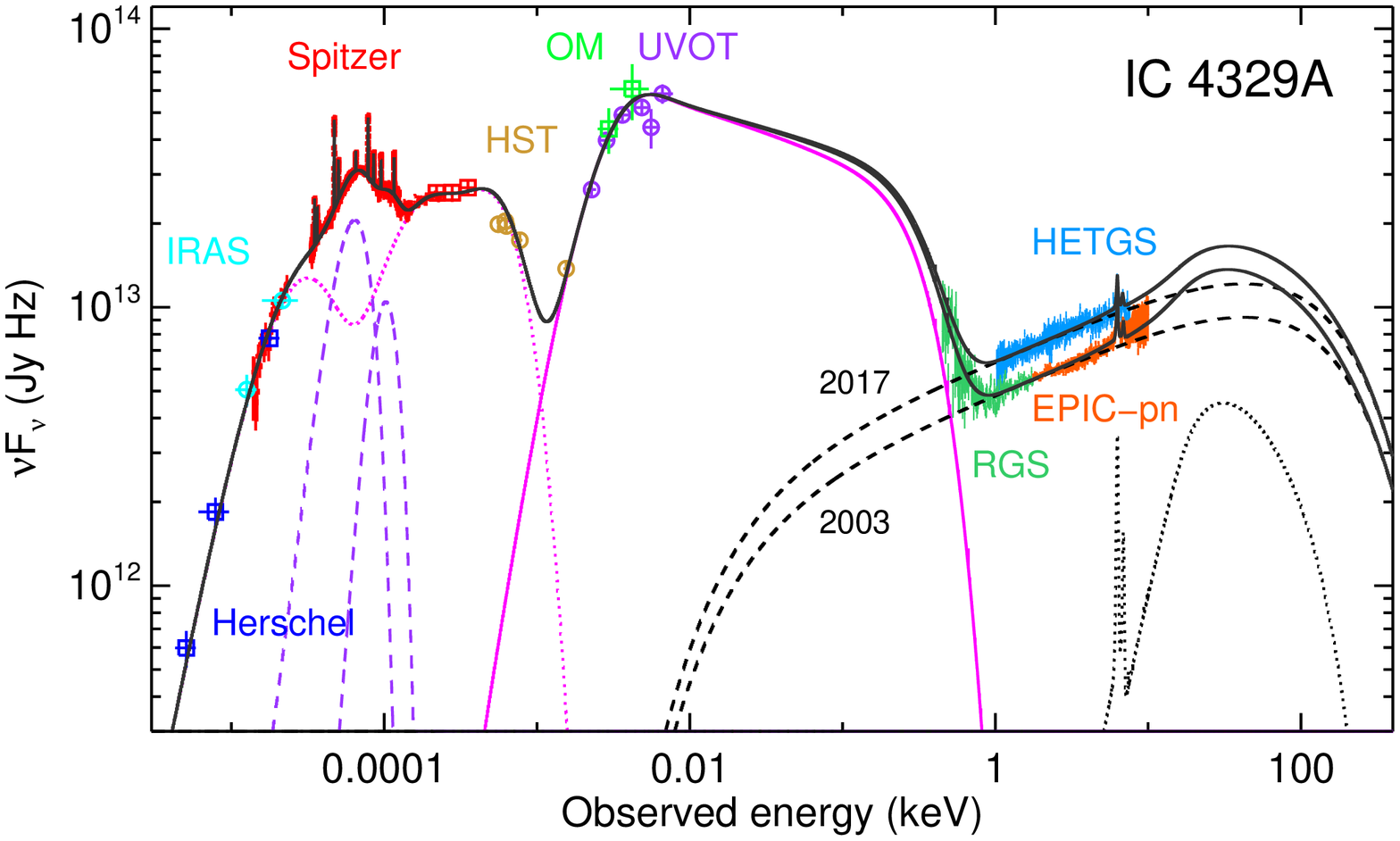}}
\caption{SED of \ic from far-IR to hard X-rays. The SED in the {\it top panel} includes the effects of reddening, X-ray absorption, host galaxy starlight emission, and the BLR and NLR emission. The SED in the {\it bottom panel} is corrected for these processes, revealing the underlying continuum. The best-fit model to the data is shown in both panels in solid black line. The contribution of individual continuum components are displayed in the {\it bottom panel}: a Comptonised disk component (solid magenta line), power-law continuum (dashed black line), X-ray reflection (dotted black line), thermal IR continuum (dotted magenta line). The model for the 9.7 and 18 \micron silicate features are also plotted (dashed purple lines). In the 2017 epoch (HETGS observation), the X-ray power-law (dashed black line) is brighter than in the 2003 epoch (\xmm observation).}
\label{SED_fig}
\end{figure}

In addition to the power-law, the soft X-ray continuum of \ic includes a `soft X-ray excess' component \citep{Capp96,Pero99,Stee05}. To model the soft excess in \ic, we use the broad-band model derived in \citet{Meh15a} for NGC~5548, in which the soft excess is modelled by warm Comptonisation (see e.g. \citealt{Mag98, Meh11, Done12, Petr13, Kubo18}). In this explanation of the soft excess, the seed disk photons are up-scattered in an optically thick and warm corona to produce the soft X-ray excess. The {\tt comt} model in \spex produces a thermal optical-UV disk component modified by warm Comptonisation, so that its high-energy tail fits the soft X-ray excess. In recent years, multi-wavelength studies have found warm Comptonisation to be a viable explanation for the soft excess in Seyfert-1 AGN (e.g. most recently in \object{Ark~120}, \citealt{Porq18}). The fitted parameters of the {\tt comt} model are its normalisation, seed photons temperature $T_{\rm seed}$, electron temperature $T_{\rm e}$, and optical depth $\tau$ of the up-scattering plasma. The primary continuum model components are shown in the SED of Fig. \ref{SED_fig}. The best-fit parameters of the power-law component, and the warm Comptonisation component (disk + soft X-ray excess), are provided in Table \ref{continuum_table}. The observed flux and intrinsic luminosity of \ic over different energy bands are given in Table \ref{luminosity_table}. 

\subsection{Reprocessed X-ray emission}
\label{reflection_sect}

\begin{figure}[!tbp]
\centering
\resizebox{1.045\hsize}{!}{\hspace{-0.45cm}\includegraphics[angle=0]{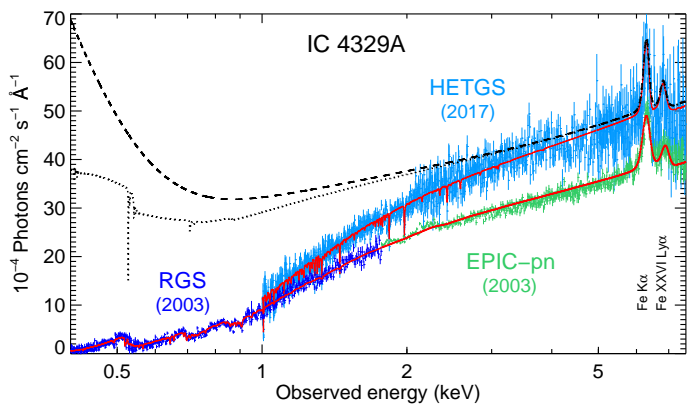}}\vspace{-0.2cm}
\caption{Overview of the X-ray spectrum of \ic taken with \xmm RGS and EPIC-pn in 2003, and \chandra HETGS in 2017. Our best-fit models to the data are also displayed (in red), with the power-law continuum being brighter in 2017. The underlying 2017 continuum model without any X-ray absorption is shown in dashed black line. For comparison, the continuum model with only absorption by the Milky Way is shown in dotted black line, which demonstrates the strong intrinsic absorption by \ic.}
\label{overview_fig}
\end{figure}

The primary X-ray continuum undergoes reprocessing, which is evident by the presence of the \FeKa line in the HETGS and EPIC-pn spectra (Fig. \ref{overview_fig}, top panel). We measure a rest line energy of ${6.39 \pm 0.01}$ keV for this line, which is consistent with neutral Fe emission. The flux of the \FeKa line is $8\pm 1 \times 10^{-13}$~\ergflux. We applied an X-ray reflection component ({\tt refl}), which reprocesses an incident power-law continuum to produce the \FeKa line and the Compton hump at hard X-rays. The {\tt refl} model in \spex computes the \FeKa line according to \citet{Zyck94}, and the Compton-reflected continuum according to \citet{Magd95}, as described in \citet{Zyck99}. The photon index $\Gamma$ of the incident power-law was set to that of the observed primary continuum ($\Gamma = 1.78$). The exponential high-energy cut-off of this incident power-law is also set to that of the observed primary power-law component at 186~keV \citep{Bren14}. In our modelling the normalisation of the incident power-law continuum is set to the average of the HETGS and \xmm observations. The ionisation parameter of {\tt refl} is set to zero to produce a cold reflection component with all abundances kept at their solar values. The reflection scale $s$ parameter of the {\tt refl} model was fitted. The {\tt refl} component was convolved with a Gaussian velocity broadening model to fit the width $\sigma_{v}$ of the \FeKa line, which is about $3400 \pm 500$~\kms. We do not require a relativistic line profile to fit the \FeKa line. 

Apart from the \FeKa line, there is another emission line at rest energy ${6.95 \pm 0.03}$ keV, which is consistent with a \ion{Fe}{xxvi} Ly$\alpha$ line (6.97 keV). We fitted this line with a simple Gaussian model. The flux of the \ion{Fe}{xxvi} line is $3\pm 1 \times 10^{-13}$~\ergflux. The line profile is unresolved with $\sigma_{v} < 2000$~\kms.

%

%

\subsection{Thermal infrared continuum}
\label{thermal_sect}

The primary AGN continuum drops from optical-UV towards lower energies. However, the observed continuum rises again from near-IR towards the mid-IR energies, and then declines again towards the far-IR energies (see Fig. \ref{SED_fig}). This is generally thought to be thermal emission from dust in AGN (e.g. \citealt{Hern16}), which mainly comes from the AGN torus. To properly fit the broad-band continuum extending from near-IR to far-IR, we require three black body ({\tt bb}) components, with different temperatures. The parameters of these components are given in Table \ref{continuum_table}. We note that we report on the 9.7 and 18 \micron dust features separately in Sect. \ref{dust_ir_sect}.

\subsection{Emission from the BLR, the NLR, and the galactic bulge}
\label{NLR_sect}

Apart from the optical and UV continuum, the photometric filters of OM and UVOT contain emission from the broad-line region (BLR) and the narrow-line region (NLR) of the AGN. Therefore, in order to correct for this contamination, we applied the emission model derived in \citet{Meh15a} for NGC~5548 as a template model for the optical and UV data of \ic. The model takes into account the Balmer continuum, the \ion{Fe}{ii} feature, and the emission lines from the BLR and NLR. We normalised this model to the intrinsic luminosity of the H$\beta$ line in \ic as derived by \citet{Marz92}, which is $2.4 \times 10^{42}$~\ergs.

To take into account the host galaxy optical and UV stellar emission in the OM and UVOT filters, we used the galactic bulge model of \citet{Kin96}, and normalised it to the \ic host galaxy flux derived from \hst images by \citet{Ben13}. This is about ${3.57 \times 10^{-15}}$ \ergcm at 5100 $\AA$.

Apart from the BLR and NLR emission in the optical and UV band, there is also infrared emission from the NLR, which is present in the \spitzer/IRS high-resolution spectrum (see Sect. \ref{dust_ir_sect}). These narrow forbidden lines include [\ion{S}{iv}] (10.511 \micron), [\ion{Ne}{ii}] (12.814 \micron), [\ion{Ne}{V}] (14.322 \micron), [\ion{Ne}{iii}] (15.555 \micron), [\ion{S}{iii}] (18.713 \micron), [\ion{Ne}{v}] (24.318 \micron), [\ion{O}{iv}] (25.913 \micron). Therefore, in our SED modelling, we take into account these components by fitting narrow Gaussian emission lines to these lines.

%
\begin{table}[!tbp]
\begin{minipage}[t]{\hsize}
\setlength{\extrarowheight}{3pt}
\caption{Best-fit parameters of the broad-band continuum model components for \ic.}
\centering
\small
\renewcommand{\footnoterule}{}
\begin{tabular}{l | c}
\hline \hline
Parameter						& Value					\\
\hline
\multicolumn{2}{c}{Primary X-ray power-law component ({\tt pow}):} 						\\
Normalisation			& ${1.77 \pm 0.01}$~(2003) \\
            			& ${2.33 \pm 0.01}$~(2017) \\
Photon index $\Gamma$		& ${1.78 \pm 0.01}$			\\
\hline
\multicolumn{2}{c}{Disk component: optical-UV and the soft X-ray excess ({\tt comt}):} 						\\
Normalisation	&		${1.4 \pm 0.1}$ \\
$T_{\rm seed}$ (eV) &	${0.8 \pm 0.1}$ \\
$T_{\rm e}$ (eV) &		${75 \pm 3}$ \\
Optical depth $\tau$ &	${33 \pm 2}$ \\
\hline
\multicolumn{2}{c}{X-ray reflection ({\tt refl}):} 						\\
Incident power-law Norm.			& ${2.05}$ (f) \\
Incident power-law $\Gamma$		& $1.78$ (f)		 			\\
Reflection scale	$s$			& $0.43 \pm 0.02$					\\
$\sigma_v$ (\kms)				& $3400 \pm 500$					\\
\hline
\multicolumn{2}{c}{Thermal IR emission components ({\tt bb}):}		 						\\
BB 1: $T$ (eV)					& $0.0077 \pm 0.0001$			\\
BB 1: Flux &  $1.59 \pm 0.02$			\\
BB 2: $T$ (eV)					& $0.042 \pm 0.001$			\\
BB 2: Flux &  $2.3 \pm 0.2$			\\
BB 3: $T$ (eV)					& $0.121 \pm 0.003$			\\
BB 3: Flux &  $3.4 \pm 0.4$			\\
\hline
\end{tabular}
\end{minipage}
\tablefoot{
The power-law normalisation of the {\tt pow} and {\tt refl} components is in units of $10^{52}$ photons~s$^{-1}$~keV$^{-1}$ at 1 keV. The normalisation of the Comptonised disk component ({\tt comt}) is in units of $10^{57}$ photons~s$^{-1}$~keV$^{-1}$. The flux of the {\tt bb} components are in $10^{-10}$ \ergflux. The high-energy exponential cut-off of the power-law for both {\tt pow} and {\tt refl} is fixed to 186~keV. The photon index $\Gamma$ of the incident power-law for the reflection component ({\tt refl}) is set to the $\Gamma$ of the observed primary power-law continuum ({\tt pow}).
}
\label{continuum_table}
\end{table}

%
\begin{table}[!tbp]
\begin{minipage}[t]{\hsize}
\setlength{\extrarowheight}{3pt}
\caption{Observed flux ($F$) and intrinsic luminosity ($L$) of \ic over various energy bands. The values correspond to the SEDs shown in Fig. \ref{SED_fig}. The continuum modelling is described in Sect. \ref{sed_sect} (Table \ref{continuum_table}).}
\centering
\small
\renewcommand{\footnoterule}{}
\begin{tabular}{l | c c}
\hline \hline
              & $F$	&  $L$		\\
Energy range  & (${10^{-11}}$ \ergflux)	&   (${10^{44}}$ \ergs)					              \\
\hline
Hard X-ray (2--10 keV) & 11.0 (2003) & 0.7 (2003) \\
                       & 14.2 (2017) & 0.9 (2017) \\
Soft X-ray (0.2--2 keV) & 3.0 (2003) & 1.4 (2003) \\
                        & 3.7 (2017) & 1.6 (2017) \\
EUV (100--1000 \AA) & ${8 \times 10^{-7}}$ & 5.8 \\
UV (1000--4000 \AA) & 0.2 & 4.5 \\
Optical (4000--7000 \AA) & 2.2 & 1.0 \\
Near-IR (0.7--3 \micron) & 23.5 & 1.5 \\
Mid-IR (3--40 \micron) & 64.6 & 3.9 \\
Far-IR (40--500 \micron) & 10.8 & 0.7 \\
\hline
1--1000 Ryd & 16.9 (2003) & 8.5 (2003) \\
            & 21.6 (2017) & 9.1 (2017) \\
Bolometric (${10^{-6}}$--${10^{3}}$ keV) & 152.0 (2003) & 24.5 (2003) \\
                                         & 164.9 (2017) & 26.4 (2017) \\
\hline
\end{tabular}
\end{minipage}
\tablefoot{The observed fluxes ($F$) include the effects of all the reddening and absorption components. The host galaxy optical and UV emission, and the AGN BLR and NLR emission, are excluded from the reported fluxes and luminosities. The IR fluxes and luminosities include all the dust emission components.}
\label{luminosity_table}
\end{table}

\section{Modelling of the ISM and the AGN wind}
\label{wind_sect}
%

\subsection{X-ray absorption by the diffuse interstellar gas}
\label{ism_sect}

Prior to modelling the ISM gas absorption in \ic, we first determined our model for the Milky Way absorption. The X-ray continuum and line absorption by the Galaxy are taken into account by applying the {\tt hot} model in \spex. This model calculates the transmission of a plasma in collisional ionisation equilibrium at a given temperature, which for neutral ISM is set to 0.5~eV. The Milky Way column density \NH is fixed to ${4.61\times 10^{20}\ \mathrm{cm}^{-2}}$\citep{Kal05} in our line of sight to \ic. The Galactic molecular \NH in our line of sight is a small fraction (17\%) of the total \NH according to \citet{Will13}.

Apart from the Milky Way absorption, an additional neutral ISM component intrinsic to the source is required for modelling the X-ray spectrum. This component is essential to fit the strong suppression of the soft X-ray continuum in \ic (see Fig. \ref{overview_fig}). Also, such a component is needed to fit the absorption features of neutral gas in the spectrum (e.g. seen in \ion{O}{i}). We thus incorporate another {\tt hot} component with its temperature fixed to 0.5~eV to fit the column density of neutral gas in \ic, which is found to be ${\NH = 3.1 \pm 0.2} \times 10^{21}$~\cm. As this is a high column density of ISM in \ic, a mildly-ionised component is also detected through primarily \ion{O}{iii}. Thus, we include another {\tt hot} component, with its temperature and \NH fitted. This mildly-ionised ISM component is found to have a temperature ${T_{\rm e} = 7.3 \pm 0.8}$~eV with ${\NH = 8 \pm 1 \times 10^{20}}$~\cm. In Sects. \ref{outflow_sect} and \ref{dust_abs_sect}, we model additional components of the absorption in \ic, produced by the AGN wind, and dust, respectively.

\subsection{X-ray absorption by the ionised AGN wind}
\label{outflow_sect}

The RGS and HETGS spectra of \ic exhibit a series of absorption lines belonging to outflowing ionised gas. We simultaneously modelled the RGS and HETGS spectra to derive a model for this ionised wind from the AGN. For photoionisation modelling and spectral fitting, we use the \pion model in \spex (see \citealt{Meh16b}), which is a self-consistent model that calculates the thermal and ionisation balance together with the spectrum of a plasma in photoionisation equilibrium (PIE). The \pion model uses the SED (Sect. \ref{sed_sect}) from the continuum model components set in \spex. During spectral fitting, as the continuum varies, the thermal and ionisation balance and the spectrum of the plasma are re-calculated at each stage. This means while using realistic broad-band continuum components to fit the data, the photoionisation is calculated accordingly by the \pion model.

To properly fit all the absorption lines from various ionic species we require three photoionisation (\pion) components with different values for the ionisation parameter $\xi$ \citep{Tar69,Kro81}. This parameter is defined as ${\xi = L / n_{\rm H}\, r^2}$, where $L$ is the luminosity of the ionising source over the 1--1000 Ryd band (13.6 eV to 13.6 keV) in \ergs, $n_{\rm H}$ the hydrogen density in cm$^{-3}$, and $r$ the distance between the photoionised gas and the ionising source in cm. The \pion components are added one at a time until all the observed features are modelled and thus our fit is no longer improved.

\begin{figure}[!tbp]
\centering
\resizebox{1.041\hsize}{!}{\hspace{-0.35cm}\includegraphics[angle=0]{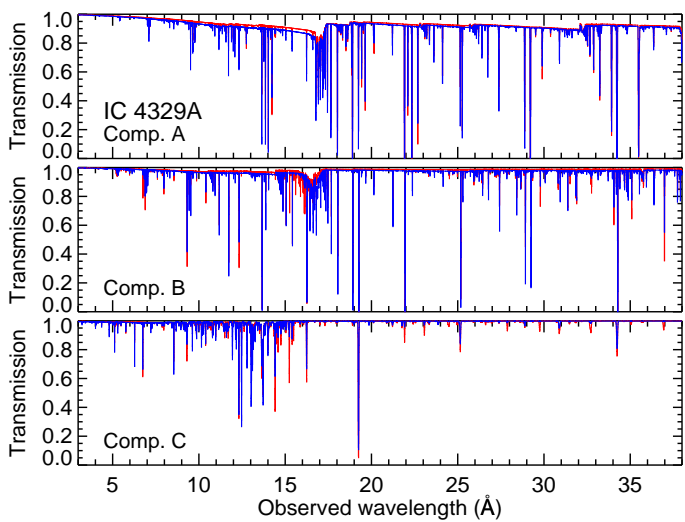}}
\caption{Model transmission spectrum of the three components of the AGN wind in \ic. Comp. A is the lowest-ionisation component, and Comp. C the highest. The model shown in blue is the one derived from the 2017 HETGS observations. The model derived from the archival 2003 \xmm observations is also shown for comparison, which is plotted in red behind the blue one.}
\label{wind_fig}
\end{figure}

According to our model, the lowest ionisation component (Comp. A) produces absorption from the Li-like ion \ion{O}{vi}, the He-like ions \ion{C}{v}, \ion{N}{vi}, and \ion{O}{vii}, and the H-like ion \ion{C}{vi}. Moreover, this component is responsible for producing a shallow unresolved transition array (UTA, \citealt{Beh01}) at about 16--17 $\AA$ from the M-shell Fe ions. The inclusion of this component improves our fit by $\Delta$\,C-stat = 820. The next ionisation component (Comp. B) primarily produces lines from the H-like \ion{N}{vii} and \ion{O}{viii}, as well as the He-like \ion{Ne}{ix} and \ion{Mg}{xi}. The lines fitted by Comp. B provide a better fit by $\Delta$\,C-stat = 440. Finally, the highest ionisation component (Comp. C) produces the H-like ions \ion{Ne}{x}, \ion{Mg}{xii}, \ion{Si}{xiv}, as well as the He-like \ion{Si}{xiii}. This component also produces the high-ionisation Fe species in the form of \ion{Fe}{xix}, \ion{Fe}{xx}, and \ion{Fe}{xxi}. The addition of the high-ionisation component further improves the fit by $\Delta$\,C-stat = 350. The column density \NH, the ionisation parameter $\xi$, the outflow velocity $v_{\rm out}$, and the turbulent velocity $\sigma_{v}$ of each \pion component are fitted. The covering fraction of all our absorption components in our modelling is fixed to unity. As the RGS and HETGS spectra are taken at different epochs, we allowed \NH and $\xi$ of the components to be different for the two epochs. The model transmission spectrum of the three \pion components (Comps. A to C) are displayed in Fig. \ref{wind_fig}. The best-fit parameters of the AGN wind components, and the ISM components, are given in Table \ref{wind_table}.

We fit all the X-ray absorption by the AGN wind in \ic with three \pion photoionisation components. However, in the X-ray analysis of \ic by \citet{Stee05}, four photoionisation components were incorporated. The three highest ionisation components in their study roughly correspond to our three \pion components, but they include an extra photoionisation component, with a very low ionisation (${\log \xi \sim - 1.37}$) and zero outflow velocity, which we do not require in our modelling. This discrepancy between the two works can be explained by differences in the modelling of the neutral X-ray absorption in \ic. The derived column density of the neutral ISM gas by \citet{Stee05} (${\NH = 1.7 \times 10^{21}}$~cm$^{-2}$) is significantly smaller than the one derived in our study (${\NH = 3.1 \times 10^{21}}$~cm$^{-2}$). However, this difference in \NH is instead modelled by the cold photoionisation component in \citet{Stee05}, whereas we associate it to the ISM absorption in \ic. Moreover, some differences in the parameterisation of the absorption components can be attributed to our inclusion of the dust model and the determination of the broad-band continuum, which result in a better fit to the spectra. Furthermore, there are extended updates and enhancements to the atomic database of \spex v3 and the new \pion photoionisation model, which were not available back in \spex v2, used by \citet{Stee05}.

%

\section{Multi-wavelength analysis of dust in \ic}
\label{dust_sect}
%

\subsection{Dust reddening in \ic}
\label{dust_red_sect}

\ic displays significant internal reddening. This is evident from both the Balmer decrement (i.e. the observed \Ha/\Hb flux ratio) and the steepness of the optical-UV continuum (see Fig. \ref{SED_fig}, top panel). The Balmer decrement of \ic has been reported in different papers to be from about 9 to 12 \citep{Wils79,Marz92,Wing96,Malk17}. This is significantly higher than the expected Balmer decrement, implying significant internal reddening. The theoretical calculations for `Case B' recombination, where gas is optically thick in Lyman lines, give \Ha/\Hb$\approx 2.85$ \citep{Bake38,Ost06}. Moreover, the observations of unreddened AGN show that the Balmer decrement of the BLR in AGN is \Ha/\Hb$\approx 2.72$ \citep{Gask17}. Importantly, the intrinsic reddening also affects the optical and UV continuum and changes the shape of the so-called `big-blue-bump' thermal emission from the accretion disk. Unlike SEDs of typical unreddened AGN (e.g. NGC~5548, \citealt{Meh15a}), the flux of \ic drops rapidly towards higher UV energies (Fig. \ref{SED_fig}, top panel), which is indicative of strong reddening.

%
\begin{table}[!tbp]
\begin{minipage}[t]{\hsize}
\setlength{\extrarowheight}{3pt}
\caption{Best-fit parameters of the final model for the AGN wind and the ISM in \ic. The parameters of the associated dust model in \ic are presented in Table \ref{dust_table}.}
\centering
\small
\renewcommand{\footnoterule}{}
\begin{tabular}{l | c}
\hline \hline
Parameter						              & Value					\\
\hline
\multicolumn{2}{c}{AGN wind Comp. A ({\tt pion}):} \\
\NH ($10^{20}$~\cm)	              &	${7 \pm 1}$ (2003) \\
                  	              &	${8 \pm 1}$ (2017) \\
$\log~\xi$ (erg~cm~s$^{-1}$)      &	${0.8 \pm 0.1}$ (2003) \\
                                  &	${1.0 \pm 0.1}$ (2017) \\
$v_{\rm out}$ (\kms)              &	${-410 \pm 30}$ \\
$\sigma_v$ (\kms)                 &	${90 \pm 20}$ \\
\hline
\multicolumn{2}{c}{AGN wind Comp. B ({\tt pion}):} \\
\NH ($10^{20}$~\cm)	              &	${6 \pm 1}$ (2003) \\
                  	              &	${6 \pm 1}$ (2017) \\
$\log~\xi$ (erg~cm~s$^{-1}$)      &	${2.1 \pm 0.1}$ (2003) \\
                                  &	${1.9 \pm 0.1}$ (2017) \\
$v_{\rm out}$ (\kms)              &	${0 \pm 20}$ \\
$\sigma_v$ (\kms)                 &	${60 \pm 10}$ \\
\hline
\multicolumn{2}{c}{AGN wind Comp. C ({\tt pion}):} \\
\NH ($10^{20}$~\cm)	              &	${11 \pm 2}$ (2003) \\
                  	              &	${22 \pm 2}$ (2017) \\
$\log~\xi$ (erg~cm~s$^{-1}$)      &	${2.79 \pm 0.07}$ (2003) \\
                                  &	${3.03 \pm 0.03}$ (2017) \\
$v_{\rm out}$ (\kms)              &	${-360 \pm 20}$ \\
$\sigma_v$ (\kms)                 &	${210 \pm 40}$ \\
\hline
\multicolumn{2}{c}{Neutral ISM component of \ic ({\tt hot}):} \\
\NH	($10^{20}$~\cm)               & ${31 \pm 1}$ \\
${T_{\rm e}}$ (eV)                & ${0.5}$ (f)	\\
\hline
\multicolumn{2}{c}{Mildly-ionised ISM component of \ic ({\tt hot}):} \\
\NH	($10^{20}$~\cm)               & ${8 \pm 1}$ \\
${T_{\rm e}}$ (eV)                & ${7.3 \pm 0.8}$	\\
\hline
\multicolumn{2}{c}{C-stat / d.o.f. = 10069 / 9280} \\
\hline
\end{tabular}
\end{minipage}
\label{wind_table}
\end{table}

Before deriving the reddening in \ic, we first fixed our model for reddening in the ISM of the Milky Way, which in our line of sight has a colour excess ${E(B-V) = 0.052}$~mag \citep{Schl11}. We applied the {\tt ebv} model in \spex to model this reddening, which incorporates the extinction curve of \citet{Car89}, including the update for near-UV given by \citet{ODo94}. The scalar specifying the ratio of total to selective extinction ${R_V = A_V/E(B-V)}$ was fixed to 3.1. 

A general model for internal reddening in AGN is lacking. Studies of reddening in AGN have found different kinds of extinction laws. For example, \citet{Hopk04} examined the SEDs of a large sample of quasars using broad-band photometry data from SDSS. They concluded that the reddening is best described by SMC-like extinction. On the other hand, the \citet{Czer04} study of composite quasar spectra from SDSS finds a `grey' (flat) extinction curve, where the extinction curve is flatter than that of the diffuse Milky Way. They also find no trace of the `2175~$\AA$ bump' that is seen in the Milky Way extinction curve. On the other hand, from a case study of a reddened AGN (\object{NGC~3227}) using HST, and comparing its spectrum with that of an unreddened AGN (NGC 4151), \citet{Cren01a} concluded that the extinction curve in the UV is even steeper than that of SMC. On the other hand, flat extinction curves have been found by \citet{Maio01,Gask04,Gask07}. 

We determined the colour excess \ebv in \ic by jointly modelling the reddening of the optical-UV continuum and the Balmer decrement. We tested different template extinction laws from \citet{Car89} (Milky Way), \citet{Czer04} (flat), and \citet{Zafa15} (steep). The user-defined multiplicative model in \spex ({\tt musr}) was used to import these reddening models into \spex. We find that a flat extinction curve \citep{Czer04} with ${\ebv = 1.0 \pm 0.1}$ is most appropriate for \ic. It gives the best-fit to the broad-band continuum with the most reasonable continuum parameters. The steeper extinction curves lead to unphysical UV-EUV continuum luminosity. This conclusion was also reached for the reddened AGN \eso \citep{Meh12}. Therefore, in our modelling of \ic, we adopt the extinction curve of \citet{Czer04}, with $R_V$ fixed to 3.1. In this case the relation between reddening and the Balmer decrement is ${\ebv = 1.475\, \log\, (R_{\rm obs} / R_{\rm int})}$, where $R_{\rm obs}$ and $R_{\rm int}$ are the observed and intrinsic \Ha/\Hb, respectively. Thus, using the reported range of the Balmer decrement in \ic (${R_{\rm obs} \approx}$ 9 to 12), and ${R_{\rm int} = 2.72}$, the reddening $\ebv$ ranges between about 0.8 and 1.0. This matches ${\ebv = 1.0 \pm 0.1}$ derived from de-reddening of the continuum (see Fig. \ref{SED_fig}). We later discuss the origin of the flat extinction curve in Sect. \ref{discussion}.

\subsection{Dust IR emission features in \ic}
\label{dust_ir_sect}

The widely known 9.7 and 18 \micron features in the mid-IR spectra of AGN belong to silicate dust (e.g. \citealt{Stur05,Henn10}). The 9.7 \micron feature is generally attributed to the stretching of the Si-O bonds in silicates, while the 18 \micron feature is attributed to O-Si-O bending in the same material (e.g. \citealt{Henn10}). These features have canonical wavelengths of 9.7 and 18 \micron in the Milky Way diffuse ISM, however, in AGN they often show deviation to longer wavelengths (e.g. \citealt{Stur05,Hatz15}). They also display much broader spectral width compared to that of the Milky Way \citep{Li08}. These features are thought to originate from silicates in the AGN dusty torus (e.g. \citealt{Pier92,Pier93,Sieb05,Maso15}). 

\begin{figure}[!tbp]
\centering
\resizebox{1.037\hsize}{!}{\hspace{-1.4cm}\includegraphics[angle=0]{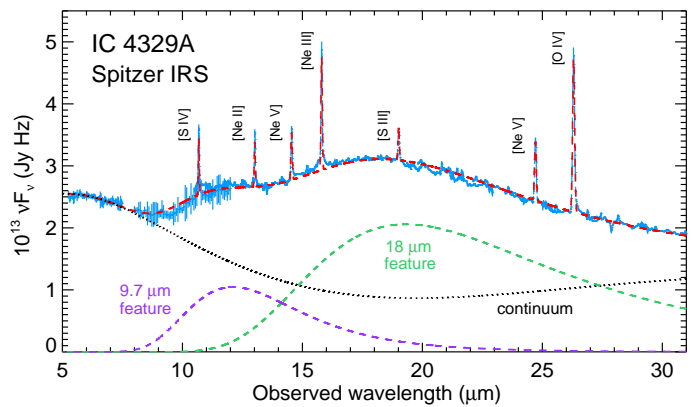}}
\caption{\spitzer IRS spectrum of \ic. The 9.7- and 18-micron emission features are shown in dashed purple and green lines, respectively. The best-fit model to the data, which includes the narrow emission lines from the AGN NLR, is shown in dashed red line. The underlying thermal IR continuum model is shown in dotted black line for comparison.}
\label{spitzer_fig}
\end{figure}

The 9.7 and 18 \micron silicate features are evident in emission in the \spitzer IRS spectrum of \ic (Fig. \ref{spitzer_fig}). The presence of these emission features in \ic were also previously reported by the AGN sample studies with \spitzer (e.g. \citealt{Gall10}). In our broad-band SED modelling, we fit these features by the means of Gaussian emission components ({\tt gaus}). This parametrisation allows us to determine the peak wavelength, flux, and the spectral width of these features. Our best-fit model to the \spitzer IRS spectrum is shown in Fig. \ref{spitzer_fig}. The results of our modelling are given in Table \ref{dust_table}, and are discussed in Sect. \ref{torus_prop_sect}.

%

\subsection{Dust X-ray absorption features in \ic}
\label{dust_abs_sect}

\begin{figure}[!tbp]
\centering
\resizebox{1.032\hsize}{!}{\hspace{-0.15cm}\includegraphics[angle=0]{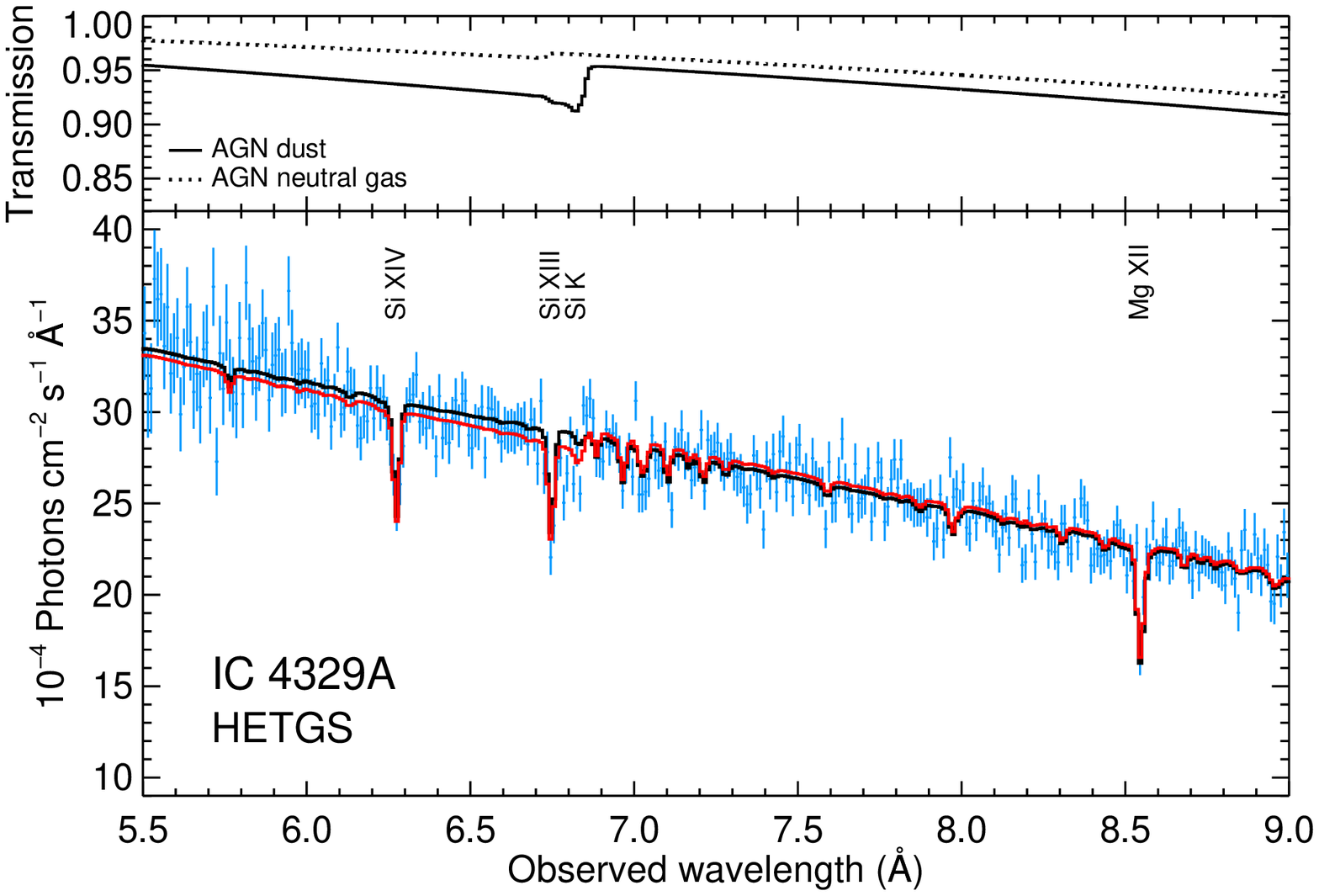}}\vspace{-0.33cm}
\resizebox{1.032\hsize}{!}{\hspace{-0.15cm}\includegraphics[angle=0]{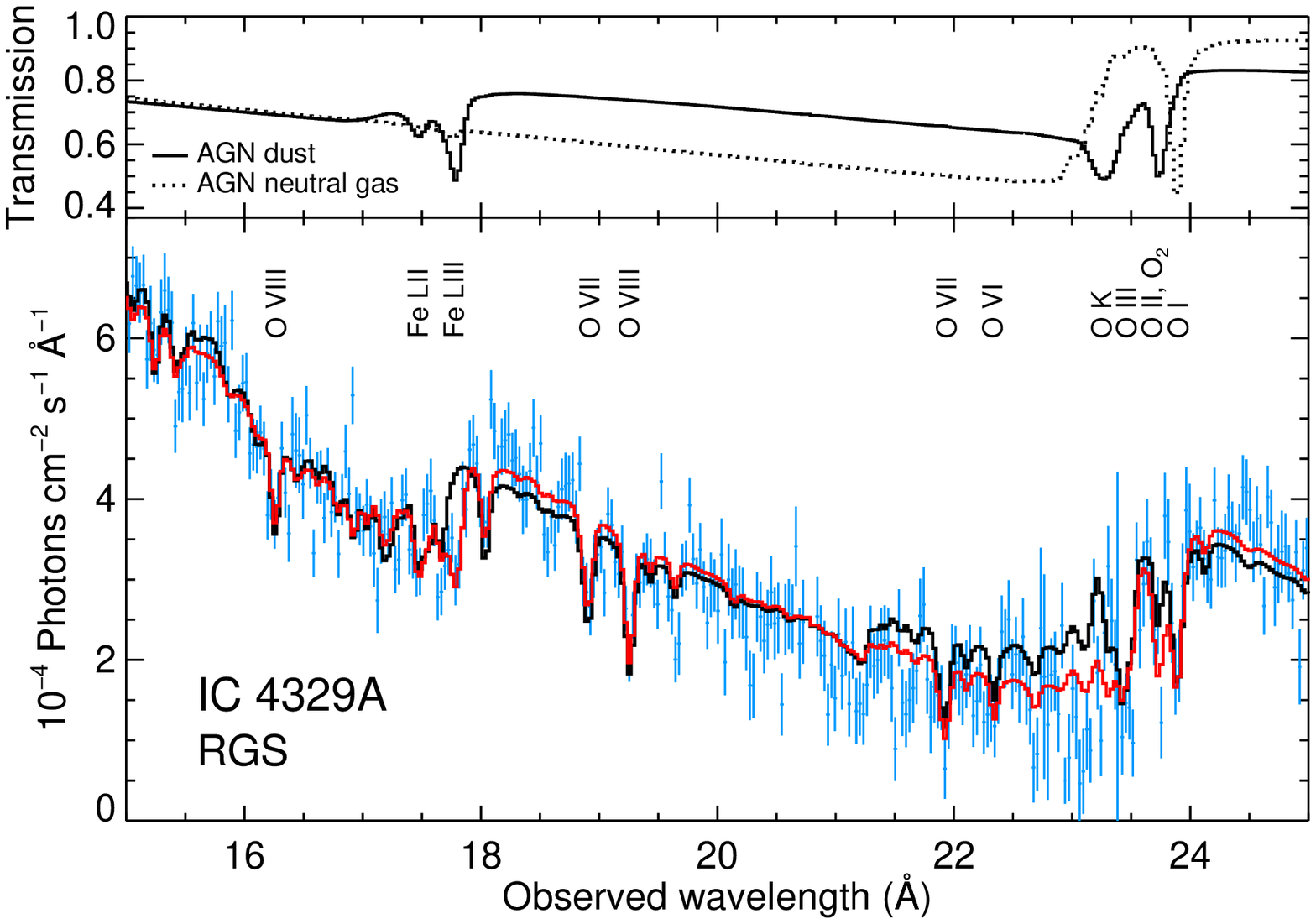}}
\caption{Stacked HETGS ({\it top panel}) and RGS ({\it bottom panel}) spectra of \ic in the K-band of Si (HETGS) and O (RGS), and the L-band of Fe (RGS). The best-fit model shown in red includes dust, whereas the one in black does not include dust. The panels above each spectrum show the transmission model by the Si, O, and Fe elements in atomic (dotted line) and dust (solid line) form in \ic. The strongest absorption features in the spectrum are labelled, including the K-edge of O and Si, and the LII- and LIII- edges of Fe from dust in the AGN, as well as lines from the ionised AGN wind. The model with dust absorption (shown in red) fits the O K and Fe LIII edges significantly better than the one without dust (shown in black).}
\label{edge_fig}
\end{figure}

A telltale sign of dust X-ray absorption is that despite atomic and ionic absorption, the photoelectric edges in the X-ray spectrum are not well fitted. This is because dust modifies the profile of the absorption edges. In the case of \ic, despite neutral and warm absorption by the ISM of the host galaxy, and the ionised gas absorption by the AGN wind, additional absorption is required at the K edge of O, the LII and LIII edges of Fe, and to lesser extent at the K edge of Si (see Fig. \ref{edge_fig}). Such stronger than expected edges are indicative of absorption by dust. The wavelength, absorption profile, and strength of these features are not consistent with atomic or ionic gas. Furthermore, the Milky Way column density and reddening in our line of sight to \ic is too low to be responsible for these features. Therefore, they can only be attributed to dust grains in our line of sight in \ic, which contain O, Si, and Fe.

%

We use the {\tt amol} model in \spex to calculate the X-ray absorption by dust (see \citealt{Pint10,Cost12}). In our modelling, we deplete the O, Si, and Fe elements from gas to dust form. For absorption in \ic the depletion factors are fitted, whereas for that of the Milky Way they are fixed. For the Milky Way contribution (albeit relatively tiny), the depletion factors are fixed to standard values taken from \citet{Jenk09}: 22\% for O, 68\% for Si, and 94\% for Fe. We apply a simple template model (hematite Fe$_2$O$_3$) to incorporate dust absorption by O and Fe by fitting the K edge of O and the LII and LIII edges of Fe. The inclusion of this dust model improves the fit to the edges significantly with {$\Delta$C-stat ${= 292}$}. While the Fe edges were fitted well at this stage, there were some remaining residuals at the O edge, which we found to be fitted well with the inclusion of molecular oxygen. This improved the fit further by {$\Delta$C-stat ${= 133}$}. Finally, we find that the fit to the Si K-edge is slightly improved by $\Delta$C-stat ${= 30}$ with the inclusion of crystal Si. We note that using AGN spectra from the existing high-resolution X-ray spectrometers, we can only constrain the column density of these elements in dust form, rather than determining the exact chemical composition of dust. Therefore, the above model represents an ad-hoc template model. Visual comparison of the best-fit models to the spectra with and without the dust model are shown in Fig. \ref{edge_fig}. The best-fit model with dust absorption (red line) fits the O K and Fe LIII edges significantly better than the one without dust (black line). The AGN wind parameters are not significantly affected by the inclusion of dust as their absorption lines are similarly fitted well regardless of the dust model. The obtained best-fit column densities of elements in dust form in \ic are given in Table \ref{dust_table}. We discuss the results in Sect. \ref{dust_comp_sect}.

%
\begin{table}[!tbp]
\begin{minipage}[t]{\hsize}
\setlength{\extrarowheight}{3pt}
\caption{Best-fit parameters of the dust model in \ic.}
\centering
\small
\renewcommand{\footnoterule}{}
\begin{tabular}{l | c}
\hline \hline
Parameter						                          & Value					\\
\hline
\multicolumn{2}{c}{Reddening component of \ic ({\tt musr}):} \\
$E(B-V)$                                      &	${1.0 \pm 0.1}$  \\
\hline
\multicolumn{2}{c}{Dust IR emission features in \ic ({\tt gaus}):} \\
\multicolumn{2}{c}{9.7 \micron feature:}  \\
$\lambda_{\rm peak}$ ($\mu$m)                 &	${13 \pm 1}$ \\
FWHM ($\mu$m)                                 &	${7 \pm 3}$ \\
Flux ($10^{-11}$ \ergflux)                    &	${5 \pm 1}$ \\
\multicolumn{2}{c}{18 \micron feature:}  \\
$\lambda_{\rm peak}$ ($\mu$m)                 &	${22 \pm 1}$ \\
FWHM ($\mu$m)                                 &	${18 \pm 1}$ \\
Flux ($10^{-11}$ \ergflux)                    &	${14 \pm 1}$ \\
\hline
\multicolumn{2}{c}{Dust X-ray absorption in \ic ({\tt amol}):} \\
$N_{\rm O-dust}$ ($10^{17}$~\cm)              & ${6.2 \pm 0.6}$ \\
$N_{\rm Si-dust}$ ($10^{17}$~\cm)             & ${1.9 \pm 0.4}$ \\
$N_{\rm Fe-dust}$ ($10^{17}$~\cm)             & ${1.5 \pm 0.1}$ \\
\hline
\end{tabular}
\end{minipage}
\tablefoot{
The AGN extinction curve of \citet{Czer04} is used to derive the reddening $E(B-V)$ with $R_{V}$ fixed at 3.1. The $\lambda_{\rm peak}$ wavelengths correspond to the rest-frame of \ic.
}
\label{dust_table}
\end{table}

\section{Discussion}
\label{discussion}
%

\subsection{Broad-band continuum of \ic}
\label{broadband_sect}

The broad-band continuum of \ic is strongly modified by internal reddening and X-ray absorption. By modelling these effects in this paper, the underlying intrinsic emission of the AGN is uncovered. While the observed SED of \ic appears significantly different from that of the archetypal Seyfert-1 galaxy NGC~5548, the underlying continuum from near-IR to hard X-rays is consistent with the global model derived for NGC~5548 in \citet{Meh15a}. The continuum from near-IR to soft X-rays can be explained by a single component, which Compton up-scatters the disk photons in an optically-thick, warm corona. The high-energy tail of this component produces the soft X-ray excess. The hard X-ray spectrum of \ic is consistent with a typical power-law (${\Gamma \approx 1.78}$), produced in an optically-thin, hot corona, which is accompanied by a neutral X-ray reflection component. The accretion-powered radiation is reprocessed into lower energies by the dusty AGN torus, which its luminosity peaks at the mid-IR band. The broad-band modelling done in this paper enables us to derive the bolometric luminosity of the AGN, which is about ${2.45 \times 10^{45}}$~\ergs in 2003, and ${2.64 \times 10^{45}}$~\ergs in 2017 (Table \ref{luminosity_table}). Taking into account the black hole mass of \ic (1--2~$\times 10^{8}$~$M_{\odot}$), these corresponds to bolometric luminosities at about 10--20\% of the Eddington luminosity. The reprocessed emission by the AGN torus accounts for about 20\% of the bolometric luminosity.

\subsection{Components of dust in \ic}
\label{dust_comp_sect}

\begin{figure}[!tbp]
\centering
\resizebox{1.015\hsize}{!}{\hspace{-0.3cm}\includegraphics[angle=0]{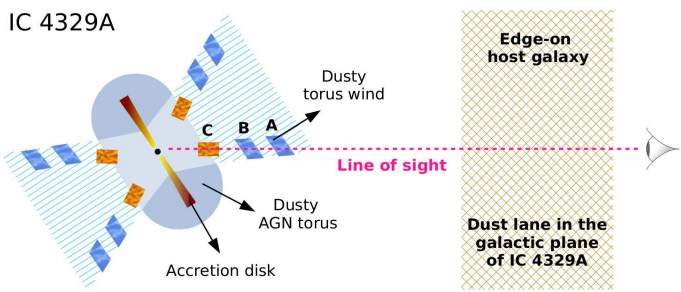}}
\caption{Illustration of our line of sight through \ic. The host galaxy and its dust lane are viewed edge-on, whereas the AGN disk and torus are tilted towards us. Our line of sight goes through the nuclear wind from the torus. The Comps. A and B of the AGN wind are located outside of the torus and are likely dusty.}
\label{cartoon_fig}
\end{figure}

In this paper we have carried out broad-band continuum modelling, together with X-ray and IR spectroscopy, to study dust in \ic. By examining our results from reddening, X-ray absorption and IR emission by dust, and comparing with previous polarisation studies, we can construct a physical picture of dust in \ic. We argue that the derived information from each of the above analyses points to the presence of two distinct components of dust in \ic: one in the ISM of the host galaxy and its associated dust lane, and the other a nuclear component in the AGN torus with likely association to the wind. In Fig. \ref{cartoon_fig} we illustrate how our line of sight towards the central engine is intercepted by these dust regions in \ic.

The relationship between the hydrogen column density \NH and the reddening \ebv can be written as ${N_{\rm{H}}\ ({\rm{cm}}^{-2}) = a \times 10^{21}\ \ebv\ (\rm{mag})}$, where $a$ is reported to have a value ranging from about 5.5 to 6.9 \citep{Bohl78,Gor75,Pre95,Guv09}. Therefore, for our derived ${\ebv = 1.0 \pm 0.1}$, the expected \NH ranges between 5.0 and ${7.6 \times 10^{21}}$~\cm. However, according to our modelling of the X-ray spectrum, the column density of neutral gas in \ic is ${\NH = 3.1 \pm 0.1 \times 10^{21}}$~cm$^{-2}$. Thus, in \ic the neutral gas alone is not sufficient to produce all the observed reddening. On the other hand, if one considers that part of the dust that causes the reddening is associated to the ionised gas in \ic (Table \ref{wind_table}), in particular the less-ionised components of the wind (i.e. a dusty warm absorber), the sum of \NH matches the expected \NH inferred from the observed reddening. The AGN wind Comps. A and B, and the neutral+warm ISM give a total \NH ${5.2 \pm 0.2 \times 10^{21}}$~\cm (from the 2003 \xmm spectra) and \NH ${5.3 \pm 0.2 \times 10^{21}}$~\cm (from the 2017 HETGS spectra). This excludes the highest ionisation component of the AGN wind. Therefore, the above results suggest that apart from the neutral ISM, the ionised gas (i.e. the warm ISM and the AGN wind) are likely dusty.

Apart from the above \NH-reddening relation, the measured column densities of dust (Table \ref{dust_table}) also indicate that some of the dust is likely associated to the AGN wind. This is mainly because the column density of Fe in dust form is too high to be feasible with Fe depletion in the ISM alone (for both the neutral and mildly-ionised components). In other words, the measured number of Fe atoms in dust form exceeds the number of Fe atoms available in gas form in the ISM. However, by considering the total \NH, including that of the AGN wind, feasible depletion factors are obtained. Taking into account the possible range of the total \NH ($4.8$ to ${5.7 \times 10^{21}}$~\cm), and the uncertainties on the column densities of O-dust, Si-dust, and Fe-dust, we derive the following depletion factors from gas to dust form in \ic: 17--23\% for O, 70--100\% for Si, and 77--98\% for Fe. These depletion factors are for the total gas in \ic, and assume the proto-solar abundances of \citet{Lod09}. The depletion factors are comparable to those in the diffuse Milky Way (e.g. \citealt{Jenk09}). We further discuss the possibility of a dusty warm absorber from the AGN torus in Sect. \ref{dust_torus_sect}.

%

\subsection{Dusty AGN torus}
\label{torus_prop_sect}

\ic displays significant optical polarisation \citep{Mart82,Wols95}. From imaging polarimetry of \ic, \citet{Wols95} find two components of polarisation. The first component is parallel to the galactic plane and the edge-on dust lane. The second component is a nuclear polarising component with a position angle approximately parallel to the galactic plane. The authors suggest that the first component arises from magnetic field in the plane of the galaxy, while the second component arises from dust scattering by an asymmetric geometry in the nucleus involving the AGN torus. Based on the polarisation of the nuclear component, \citet{Wols95} argue that the torus must be significantly tilted in our line of sight, while we can still see the central engine. In this case, \ic is still perceived spectroscopically as a type-1 AGN, but the inner edge of the torus will be viewed as an elliptical ring which introduces an asymmetry into the scattering geometry.

The 9.7 and 18 \micron silicate features are commonly seen in absorption in type-2 AGN, and in emission in type-1 AGN. This is because of strong silicate absorption by the AGN torus in type-2 AGN, which obscures the line of sight. However, in \ic, our view of the accretion disk is only partially obscured by the torus. Thus, silicate emission from the inner walls of the torus dominates over any weaker silicate absorption in our line of sight. Furthermore, highly-inclined (i.e. edge-on) galaxies often show the 9.7 and 18 \micron features in absorption, rather than emission (e.g. \citealt{Alon11}). This can be attributed to silicate absorption in the disk of the galaxy. However, in \ic the 9.7 and 18 \micron features are seen in emission, rather than absorption \citep{Deo09,Alon11}, which is also found by our analysis of the \spitzer/IRS spectrum (Fig. \ref{spitzer_fig}). As suggested by \citet{Deo09}, this is likely because our line of sight does not intersect dense dust clouds in the ISM of the galaxy. 

\citet{Tris09} have carried out a survey with MID-infrared Interferometric instrument (MIDI) at the VLTI on 13 bright AGN (including \ic). This is to resolve the nuclear dust emission in these AGN and study their spatial distribution. However, they find that \ic emission is unresolved, which implies that the bulk of the mid-IR emission (12~\micron) is concentrated at ${< 10.8}$~pc \citep{Tris09}. This suggests that mid-IR dust emission from the nucleus (i.e. the dusty torus) dominates over non-nuclear emission (i.e. the dust lane).

From our modelling of internal reddening in \ic (Sect. \ref{dust_red_sect}), we derived ${E(B-V) \approx 1.0}$. The reddening is best described with a grey extinction curve \citep{Czer04}, which is relatively flat in the UV with no steep rise into the far-UV. This implies that dust in \ic is different from that of the Milky Way. From the \spitzer study of dust features in a sample of AGN, \citet{Xie17} conclude that all sources require that the dust grains are micron-sized (typically ${\sim 1.5}$ micron), which is much larger than the sub-micron sized Galactic interstellar grains. This would imply a grey (flat) extinction law for AGN \citep{Xie17}, which is what we find for \ic from our SED modelling. Therefore, the torus dust grains in \ic are likely larger than those in the diffuse Milky Way.

In \ic, we find that the 9.7 and 18 \micron silicate emission features peak at wavelengths higher than the standard ISM silicate dust (Table \ref{dust_table}). A recent census of the silicate features in the mid-IR spectra of AGN by \citet{Hatz15} shows that $\lambda_{\rm peak}$ of the 9.7 \micron emission feature (in the rest frame) is often shifted to longer wavelengths relative to the nominal wavelength of 9.7 \micron. This shift is much less present when the feature is in absorption. \citet{Hatz15} report that the shift nearly always occurs in AGN dominated spectra, where the fractional contribution of the AGN component to the total luminosity between 5 and 15 \micron is $> 0.7$. From our modelling we find this is indeed the case for \ic, as this fraction is $\approx 0.80$. The peak wavelength, spectral width, and relative strength of the 9.7 and 18 \micron features are thought to depend on the grain composition and size (e.g. \citealt{Xie17}). For example, amorphous olivine gives the longest peak wavelength for the 9.7 \micron feature and the highest ratio of the 18 \micron feature to the 9.7 \micron feature \citep{Xie17}, which is the case for \ic. Moreover, the shifting and broadening of the silicate features have been explained by \citet{Li08} using porous composite dust, consisting of amorphous silicate. They suggest that such porous dust is expected in the dense circumnuclear region of the AGN, as a consequence of grain coagulation.

\subsection{Dusty wind from the AGN torus}
\label{dust_torus_sect}

From our photoionisation modelling of the AGN wind in \ic (Sect. \ref{wind_sect}), we find that it consists of three ionisation components. According to our modelling, the two lowest ionisation components (Comps. A and B) are consistent with having no significant variability between the 2003 and 2017 epochs. Their column density \NH and ionisation parameter $\xi$ remain unchanged (Table \ref{wind_table}). However, the highest ionisation component (Comp. C) displays significant variability in \NH and $\xi$. This increase in $\xi$ of Comp. C between the two epochs matches the increase in the luminosity of the X-ray continuum (Fig. \ref{SED_fig}, Tables \ref{continuum_table} and \ref{luminosity_table}), suggesting that Comp. C has responded to the variability of the ionising SED. Therefore, the recombination timescale ($t_{\rm rec}$) of Comp. C needs to be shorter than the spacing between the two epochs, while the lack of variability in Comps. A and B suggests that their $t_{\rm rec}$ are longer than this limit. From our photoionisation modelling, the product $t_{\rm rec} \times n_{\rm H}$ for each component is yielded. We find $t_{\rm rec}$ is about ${60~n_{4}^{-1}}$ days for Comp. A, ${49~n_{4}^{-1}}$ days for Comp. B, and ${9~n_{4}^{-1}}$ days for Comp. C, where ${n_{4}}$ represents ${n_{\rm H} = 10^{4}}$~cm$^{-3}$. Hence, from the limit on $t_{\rm rec}$, based on the non-variability of Comps. A and B, and variability of Comp. C, we can place a limit on the density $n_{\rm H}$ of each component. Consequently, from this $n_{\rm H}$ limit and the definition of the ionisation parameter (${\xi = L / n_{\rm H}\, r^{2}}$), limit on the location $r$ of each component from the ionising source is computed. We find Comp. A is at ${r > 350}$~pc, Comp. B at ${r > 83}$~pc, and Comp. C at ${r < 93}$~pc. For comparison, the inner radius of the AGN torus in \ic, derived by the torus modelling of \citet{Alon11} is 2.7~pc. 

The observational characteristics of the wind in \ic match the warm-absorber type winds in type-1 AGN. Such winds are most consistent with an origin from the AGN torus (e.g. \citealt{Kaas12,Meh18}). The location of the wind components in \ic (Comps. A and B), and their relatively low outflow velocities, are consistent with a torus wind. As discussed in Sect. \ref{dust_comp_sect}, a component of dust in \ic arises from the AGN torus. Thus, the reddening and X-ray absorption from this nuclear component in our line of sight is most likely associated to a wind from the torus. The two lowest ionisation components (Comps. A and B) of the AGN wind are most likely dusty. These components, which are located outside of the dusty torus, are beyond the dust sublimation radius, and therefore they can host dust. Interestingly, there are observational evidence of extended mid-IR emission from dust along the polar directions in type-2 AGN, such as the \object{Circinus} galaxy \citep{Tris14}, which may be produced by the dusty outflows from the AGN torus.

Currently using \xmm and \chandra, the dust X-ray features are only detectable in a very few X-ray bright AGN. The proposed {\it Arcus} mission \citep{Smit16} with its unprecedented sensitivity and energy resolution in the soft X-ray band would provide a major breakthrough in the X-ray spectroscopy of dust. It would enable us to distinguish between different chemical compositions of dust in AGN, providing important insight into the evolutionary path of AGN. Furthermore, the upcoming \athena observatory \citep{Nand13} extends the wind studies to a larger population of AGN, from different types and ages, which is key for understanding the role and impact of AGN winds in galaxy evolution.

\section{Conclusions}
\label{conclusions}

In this paper we have carried out broad-band continuum modelling spanning far-IR to hard X-rays, combined with high-resolution X-ray and IR spectroscopy, to investigate the nature and origin of dust in \ic. From the findings of our investigation we conclude the following. 

\begin{enumerate}
\item There are two distinct components of dust in \ic: an ISM dust lane component, and a nuclear component. The nuclear dust component originates from the AGN torus and its associated wind. Dust in the AGN torus is seen through IR emission, while the dust in the torus wind is detected through reddening and X-ray absorption in our line of sight.  
\medskip
\item The AGN wind in \ic consists of three ionisation components. They have moderate outflow velocities: $-340$ to $-440$ \kms for two of the components, while the other is consistent with zero outflow velocity.
\medskip
\item According to our variability analysis, the two lowest ionisation components of the AGN wind show no long-term changes between historical and new observations (${\sim 14}$ years apart), while the highest component shows changes in \NH and ionisation parameter $\xi$. From the recombination timescale analysis, we derive limits on the distance of the wind components from the central engine. The two lowest ionisation components are at ${r > 350}$~pc, and ${r > 83}$~pc, while the highest ionisation component is at ${r < 93}$~pc.
\medskip
\item The total internal reddening in \ic is $\ebv \approx 1.0$. The reddening is most consistent with a grey (flat) extinction law. 
\medskip
\item From high-resolution X-ray spectroscopy of dust in \ic we derive the total depletion factors from gas into dust for O, Si, and Fe. They correspond to 17--23\% for O, 70--100\% for Si, and 77--98\% for Fe.
\medskip
\item The dust grains associated to the AGN torus and its wind in \ic are likely larger than the Milky Way ISM dust, and are in a porous composite form, containing amorphous silicate with iron and oxygen.
\medskip
\item From on our modelling of the continuum from far-infrared to X-rays, we derive a bolometric luminosity of about ${2.5 \times 10^{45}}$~\ergs for \ic, which corresponds to 10--20\% of the Eddington luminosity. 
\end{enumerate}

\begin{acknowledgements}
M.M. and E.C. are supported by the Netherlands Organisation for Scientific Research (NWO) through The Innovational Research Incentives Scheme Vidi grant 639.042.525. This research has made use of data obtained from the Chandra Data Archive, and software provided by the Chandra X-ray Center (CXC). This work made use of data supplied by the UK \swift Science Data Centre at the University of Leicester. This work is based on observations made with the NASA/ESA Hubble Space Telescope, obtained from the data archive at the Space Telescope Science Institute. This research has made use of the NASA/IPAC Infrared Science Archive, which is operated by the Jet Propulsion Laboratory, California Institute of Technology, under contract with the National Aeronautics and Space Administration. We thank Fred Lahuis for help with the reduction of the \spitzer IRS spectra, and Michiel Min for useful discussions. We thank the anonymous referee for the useful comments.
\end{acknowledgements}

\end{document}